\def\vortexrootdriver{1}
\makeatletter

\@ifundefined{showhyphens }{}{%
  \expandafter\def\csname showhyphens \endcsname#1{%
    \setbox0\vbox{%
      \color@begingroup
      \everypar{}\parfillskip\z@skip\hsize\maxdimen\normalfont
      \pretolerance\m@ne\tolerance\m@ne\hbadness\z@\showboxdepth\z@\ #1%
      \color@endgroup}}%
}

\makeatother

\ifdefined\vortexrootdriver\else
  \errmessage{Do not compile profiles/arxiv/main.tex directly. Compile vortex.tex from the repository root instead.}%
  \expandafter\endinput
\fi
\documentclass[letterpaper,twocolumn,10pt]{article}
\usepackage{style/usenix/usenix-2020-09}

\newcommand{\VortexSystemName}{Vortex}
\newcommand{\VortexHashName}{VortexHash}
\newcommand{\VortexOverlayName}{VDHT}
\newcommand{\VortexDatasetAppendixRef}{the online appendix~\cite{VortexOnlineAppendix}}
\newcommand{\VortexTestbedLatencySubject}{\VortexOverlayName}
\usepackage{amsmath}
\usepackage{graphicx}
\usepackage{booktabs}
\usepackage{array}
\usepackage{tabularx}
\makeatletter
\@ifclassloaded{acmart}{}{%
  \usepackage{amssymb}
}
\makeatother
\providecommand{\Description}[2][]{}

\newcommand{\VortexTighten}[1]{\par\vskip#1\relax}

\DeclareRobustCommand{\VortexAppendixRef}[1]{%
  \ifvortexappendixrefsonline
    the online appendix~\cite{VortexOnlineAppendix}%
  \else
    Appendix~\ref{#1}%
  \fi
}
\DeclareRobustCommand{\VortexAppendixRefCap}[1]{%
  \ifvortexappendixrefsonline
    The online appendix~\cite{VortexOnlineAppendix}%
  \else
    Appendix~\ref{#1}%
  \fi
}
\newcolumntype{L}{>{\raggedright\arraybackslash}X}

\graphicspath{{figures/}}
\makeatletter
\@ifpackageloaded{microtype}{%
  \microtypecontext{spacing=nonfrench}%
}{}
\makeatother
\makeatletter
\@ifclassloaded{acmart}{%
  \setlength{\headheight}{15.62549pt}
}{}
\makeatother

\newcommand{\vortexpapertitleplain}{Vortex: Efficient Decentralized Vector Overlay for Similarity Search and Delivery}
\newcommand{\vortexpapertitledisplay}{Vortex: Efficient Decentralized Vector Overlay\\ for Similarity Search and Delivery}
\newcommand{\vortexpapertitlewrapped}{\texorpdfstring{\vortexpapertitledisplay}{\vortexpapertitleplain}}

\newif\ifvortexincludeacknowledgments
\vortexincludeacknowledgmentsfalse

\newif\ifvortexincludeappendix
\vortexincludeappendixtrue

\newif\ifvortexappendixrefsonline
\vortexappendixrefsonlinefalse

\renewcommand{\VortexSystemName}{Semord}
\renewcommand{\VortexHashName}{VHash}
\renewcommand{\VortexOverlayName}{VecDHT}
\renewcommand{\vortexpapertitleplain}{Semord: Learned Semantic-Preserving Placement and Low-Fanout Routing for Distributed Vector Search}
\renewcommand{\vortexpapertitledisplay}{Semord: Learned Semantic-Preserving Placement\\ and Low-Fanout Routing for Distributed Vector Search}

\title{\vortexpapertitlewrapped}
\date{}
\author{%
  {\rm Shengze Wang\textsuperscript{1}, Yi Liu\textsuperscript{1}, Yifan Hua\textsuperscript{1},
  Xiaoxue Zhang\textsuperscript{2}, Chen Qian\textsuperscript{1}}\\
  \textsuperscript{1}University of California, Santa Cruz\\
  \textsuperscript{2}University of Nevada, Reno
}
\hypersetup{%
  pdftitle={\vortexpapertitleplain},
  pdfauthor={Shengze Wang, Yi Liu, Yifan Hua, Xiaoxue Zhang, Chen Qian}
}

\vortexincludeacknowledgmentstrue
\vortexincludeappendixtrue
\vortexappendixrefsonlinefalse
\renewcommand{\VortexDatasetAppendixRef}{\VortexAppendixRef{app:scale-details}}
\renewcommand{\VortexTestbedLatencySubject}{\VortexSystemName}

\newcommand{\VortexPrintBibliography}{%
  \begingroup
  \sloppy
  \hbadness=10000\relax
  \emergencystretch=3em\relax
  \bibliographystyle{plain}
  \bibliography{references}
  \endgroup
}

\begin{document}

  \maketitle
  \begin{abstract}
  Vector databases are increasingly deployed in distributed settings where different users, sites, or domains maintain vector data.
Existing vector databases rely on a coordinator to record which shards store which parts of the vector space and to route each query to those shards. In a decentralized setting, peers may join, leave, or move data without a trusted node tracking every change, and outdated routing information can therefore send queries to the wrong peers or require contacting many peers, reducing vector retrieval recall and increasing network latency.
We present \textbf{\VortexSystemName}, a decentralized vector search overlay system that achieves high recall by routing each ANN query to a small set of relevant peers, without relying on a centralized coordinator.
\VortexSystemName{} addresses this problem by making semantic locality \emph{routable}: 1) We propose \VortexHashName{} to place semantically realted vectors near each other in the overlay key space while avoiding load imbalance, so that each query only needs to contact a small neighborhood of peers for distributed local ANN ranking. 2) We design \VortexOverlayName, a communication protocol that maintains decentralized routing, region metadata, churn resilience, and \VortexHashName{} updates under membership and workload changes.
Our extensive experiments on real testbed show that \VortexSystemName{} improves recall by more than 15\% and reduces contacted peers by over 60\% compared with decentralized baselines. \VortexSystemName{} also approaches the recall and latency of a centralized oracle baseline while reducing peak peer-local ANN index memory by more than 2$\times$. Controlled large-scale simulations further  show that \VortexSystemName{} scales across real-world embedding workloads and remains robust under churn for scoped vector retrieval as a decentralized overlay.

  \end{abstract}

\section{Introduction}
\label{sec:introduction}

Persistent retrieval systems are becoming the core infrastructure for LLM and Agentic AI applications. Modern agent frameworks increasingly treat retrieval as a long-term memory substrate: short-lived state remains thread-scoped, while durable memories are stored externally and recalled across conversations, users, and tasks~\cite{Lewis2020RAG, Park2023GenerativeAgents}. At the same time, vector databases have emerged as an important class of persistent retrieval systems~\cite{Pan2024VDBMSSurvey}. By storing vector representations and retrieving semantically relevant records on demand, they provide a practical substrate for long-term memory. As retrieval increasingly lies on the latency-critical path of user queries, the systems challenge is how that memory can be stored, organized, and retrieved efficiently under realistic deployment constraints.

We focus on \emph{scoped, dynamic} vector data collections. Retrieval is increasingly bounded by user, project, or workspace, and the underlying vectors evolve continuously as new memories are added. We use \emph{scope} to denote this collection-level eligibility constraint. When eligible vectors are distributed across peers, approximate top-$k$ retrieval requires \emph{peer selection} as well as local approximate nearest-neighbor (ANN) search: the system must determine which peers are most likely to hold useful results. In the \emph{weakly coordinated} setting we target, queries may enter at any participating peer, and peer selection cannot depend on a permanent query router or a globally up-to-date shard directory.

Existing high-performance vector systems scale effectively, but they optimize a different point in the design. Managed vector services and distributed vector databases typically rely on centralized query routing: centralized data planes that maintain placement and selection metadata and determine which shards or workers should be searched. Milvus exemplifies this coordinated design point~\cite{Wang2021Milvus}. VBase further shows that, even within a single administrative domain, supporting vector retrieval with richer query semantics requires specialized system abstractions~\cite{Zhang2023VBase}. Recent serverless vector systems follow the similar design pattern: Vexless introduces a global coordinator to assign work to cloud-function instances based on available resources~\cite{Su2024Vexless}. Such designs are effective when centralized coordination is available and acceptable. They are less suitable, however, for weakly coordinated deployments in which data is distributed across peers and query routing cannot depend on a single, continuously available coordinator.

Without an effective peer-selection signal, decentralized search must fall back to broad fan-out: probing many peers and merging their results. This scatter--gather approach increases query-path traffic, contacts more peers, and amplifies tail latency. As the system scales, these costs can make decentralized retrieval inefficient under impractical communication budgets.

This paper targets a missing design point: \emph{high-recall semantic retrieval under weak coordination, without network-wide control}. Classical decentralized overlays such as DHTs randomize key placement to balance load and support efficient exact-key lookup~\cite{Stoica2001Chord}, but such placement can not preserve semantic locality among vectors. Earlier semantic-overlay systems recognized that content-aware placement can improve peer-to-peer search~\cite{Tang2003PSearch,Li2004SSW}. \VortexSystemName{} builds on this premise: \emph{semantic locality can guide overlay routing}. However, locality-aware placement alone does not tell which ingress peer currently owns the target regions of the vector space.

We present \emph{\VortexSystemName}, a decentralized overlay system for vector retrieval. \VortexSystemName{} is built around a simple principle: \emph{semantic locality must become routable}. Under weak coordination, the global organization of vectors must simultaneously provide \emph{routability}, preserve \emph{semantic locality}, and remain balanced and maintainable under skew and churn. Conventional hash-based overlays provide efficient routing and randomized load distribution, but deliberately destroy semantic locality. Locality-preserving partitions retain semantic structure, but do not by themselves provide a routable overlay that can be maintained as membership and data evolve.

\VortexSystemName{} bridges this gap with \emph{\VortexHashName}, a routing-aware, load-balanced, semantic-preserving learned hash function that maps semantically related embeddings into an ordered overlay key space. Nearby key regions therefore become meaningful routing signals rather than arbitrary hash buckets, while capacity is allocated across the key space to mitigate hotspots caused by semantic skew. An ingress peer hashes a query into this space and routes toward a small set of candidate peers most likely to hold relevant vectors.

\VortexSystemName{} separates \emph{global peer selection} from \emph{local ranking}. Decentralized semantic expansion and bounded repair compensate for residual placement and routing error, while each contacted peer uses its local ANN index to retrieve and rank candidate vectors. A co-designed \emph{\VortexOverlayName} protocol provides decentralized routing, region and ownership discovery, repair, and \VortexHashName{} state maintenance, allowing peer selection to remain selective as membership changes without relying on a permanent coordinator. \VortexSystemName’s core insight is that decentralized similarity search scales only when semantic locality is compiled into overlay locality, so that the overlay serves not merely as a substrate for ownership and exact lookup, but as a substrate for search.

This paper makes four main contributions.

First, we identify a missing systems design point for modern vector retrieval: \emph{high-recall ANN search over scope-eligible vectors under weak coordination}. Existing approaches rely on centralized routing to direct queries. We realize selective decentralized retrieval through making semantic locality directly routable.

Second, we introduce \emph{\VortexHashName}, a routing-aware, semantic-preserving, load-balanced learned hash function for vector retrieval. \VortexHashName{} maps semantically related vectors to nearby regions of an ordered overlay key space while allocating capacity according to semantic density. This makes the resulting key space useful for both selective routing and load balancing under skew.

Third, we design \emph{\VortexOverlayName}, a decentralized routing and maintenance protocol co-designed with \VortexHashName{} to make its learned semantic key space usable under weak coordination. \VortexOverlayName{} provides decentralized owner and region discovery, routing, repair, and learned-state maintenance under membership change, enabling bounded-fanout peer selection without a permanent coordinator. It further integrates bounded repair and decentralized semantic expansion with peer-local ANN indexes for final candidate retrieval and ranking.

Fourth, we evaluate \VortexSystemName{} as an end-to-end networked retrieval system. 
Across a real-world testbed and larger-scale controlled experiments, \VortexSystemName{} achieves higher recall while contacting over 60\% fewer peers than competitive decentralized baselines. It also approaches the recall and latency of a centralized oracle baseline while reducing peak peer-local ANN-index memory by more than 2$\times$.

\VortexTighten{-4ex}
\section{Background and Motivation}
\label{sec:background}

\noindent\textbf{Decentralization as a deployment constraint.}
\VortexSystemName{} is motivated by a shift in where now retrieval state resides. In emerging deployments, relevant vectors are distributed across the peers. Persistent agent memory is one example: modern agent architectures increasingly externalize long-lived state and retrieve it dynamically during planning and interaction.

Distributed AI infrastructures reinforce this trend. AI-RAN systems increasingly place inference on radio infrastructure at the edge~\cite{NVIDIA_AIRAN_Glossary_2026,NVIDIASoftBank2024AIRAN}, while recent work on AI-native 6G RAN considers distributed AI execution across far-edge, near-edge, and cloud sites~\cite{Ananthanarayanan2025DistributedAIRAN}. Satellite--ground platforms such as LEOEdge similarly study inference across resource-heterogeneous and mobile nodes~\cite{Yao2025LEOEdge}, and edge-device RAG systems already perform retrieval under tight local memory and compute constraints~\cite{Seemakhupt2024EdgeRAG}. Across these settings, retrieval is increasingly \emph{scoped}, \emph{dynamic}, and \emph{latency-sensitive}, while the underlying state is distributed across independently evolving sites. This makes it increasingly difficult---and in some deployments undesirable---to place every query behind a single, permanently available and fully trusted routing service.

Also, recent work highlights privacy risks that favor localized storage and limited exposure~\cite{Park2023GenerativeAgents,Wang2025AgentMemoryPrivacy}. Federated and multi-repository RAG exhibit a similar structure, with knowledge distributed across repositories or administrative domains and selective source retrieval emerging as a practical systems problem~\cite{Guerraoui2025RAGRoute}.

\smallskip
\noindent\textbf{Centralized systems and local ANN are not enough.}
This does not make coordinated vector systems obsolete; rather, it exposes a different design point. Systems such as Milvus, VBase, and Vexless scale effectively when a control plane can maintain global metadata and decide where queries should execute~\cite{Wang2021Milvus,Zhang2023VBase,Su2024Vexless}. In parallel, systems such as LANNS and SPANN have substantially improved local ANN indexing and execution, including index construction, memory efficiency, and storage-aware search at large scale~\cite{Doshi2021LANNS,Chen2021SPANN}. These advances address complementary parts of the retrieval stack, but not the global routing problem posed by weak coordination: given a scoped query and no permanently available trusted router, which small set of peers should execute the local search?

This distinction also exposes a second systems pressure. High-performance vector systems can achieve low latency at scale by provisioning substantial memory-resident index state or by introducing more sophisticated memory--storage and computation hierarchies~\cite{Doshi2021LANNS,Chen2021SPANN}. \VortexSystemName{} does not replace these local ANN techniques. Instead, it targets the complementary problem they leave unresolved: \emph{selective peer selection} when eligible vectors are distributed and no permanent coordinator maintains a globally current view of their placement.

\smallskip
\noindent\textbf{The missing design point.}
A decentralized retrieval substrate must satisfy three objectives simultaneously: preserve sufficient \emph{semantic locality} to concentrate true neighbors, expose enough \emph{routable structure} for the overlay to exploit key order directly, and remain \emph{robust under skew and churn} so that dense regions do not become persistent hotspots and routing remains effective as membership evolves. It must also bound query fanout explicitly, so that retrieval does not dominate end-to-end communication and latency. 
\section{\VortexSystemName{} Design}
\label{sec:design}

\begin{figure*}[htbp]
    \centering
    \VortexTighten{-4ex}
    \includegraphics[width=\linewidth]{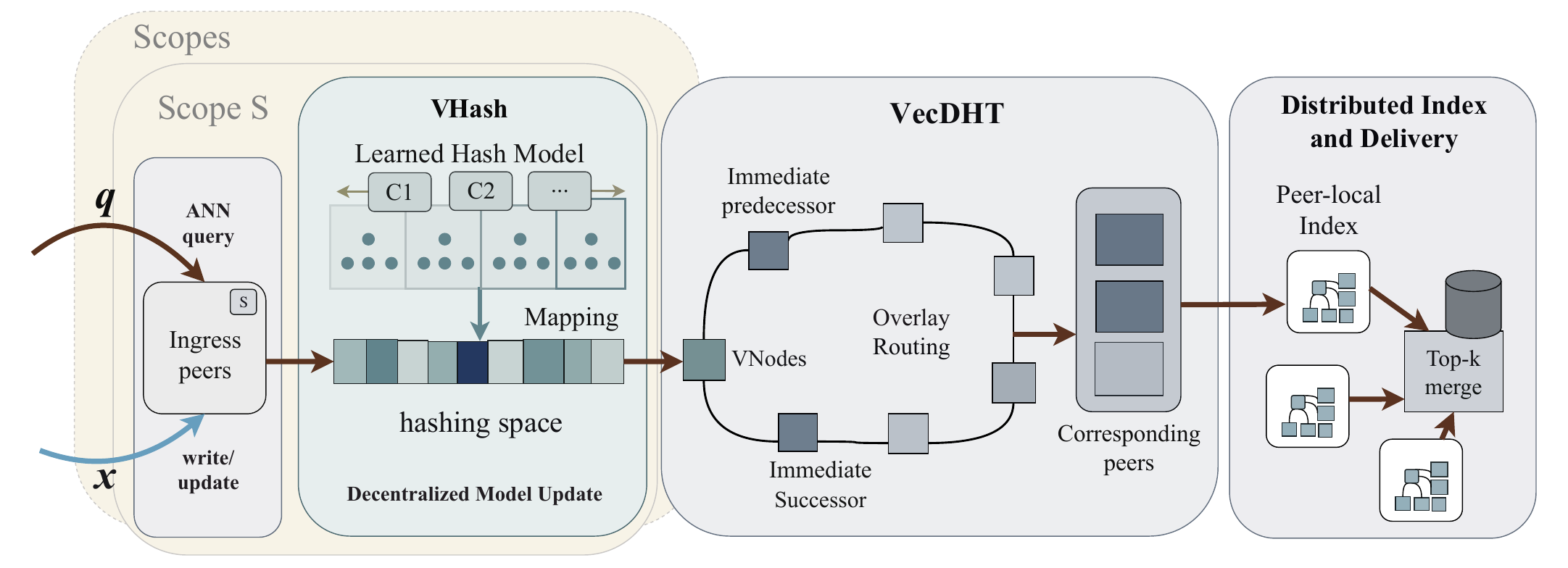}
    \Description{System overview showing query ingress, \VortexHashName{} semantic key mapping, \VortexOverlayName{} overlay routing and directory state, bounded peer selection, peer-local ANN search, and payload retrieval across scoped peers.}
    \VortexTighten{-2ex}
    \caption{Overview of \VortexSystemName.}
    \VortexTighten{-2ex}
    \label{fig:vortex-overview}
\end{figure*}

Figure~\ref{fig:vortex-overview} summarizes the operating picture. \VortexSystemName{} solves a routing
problem before it solves an ANN problem. In a weakly coordinated deployment, the hard part
of top-$k$ vector retrieval is not only comparing embeddings, but first identifying which
eligible peers are worth contacting at all. \VortexSystemName{} addresses this by making semantic
locality \emph{routable}. For each scope, \VortexHashName{} maps vectors into an ordered $b$-bit
key space whose neighborhoods preserve coarse semantic locality and whose interval widths
reflect semantic density. \VortexOverlayName{} then maintains live distributed ownership and region
directory state over that same key space under joins, leaves, and failures. Peer-local ANN
indexes perform the final ranking only at the peers surfaced by these first two layers.
The result is a two-stage retrieval pipeline: overlay-level peer selection followed by
peer-local ranking.

For clarity, we fix one scope $S$ and omit the scope superscript below; each scope
instantiates its own \VortexHashName{} namespace over the same implementation
substrate. In practice, every key, metadata token, and region identifier is
scope-qualified, so routing never leaves the eligible collection. 

Within a scope, \VortexSystemName{} must satisfy four requirements. It must preserve enough semantic
locality that nearby key ranges concentrate promising vectors. It must keep fanout
explicitly bounded even when learned placement is imperfect. It must remain balanced under
density skew. And it must continue operating without a globally current shard map or a
permanent trusted router. \VortexSystemName{} meets these requirements with three co-designed
components. \emph{\VortexHashName} is a learned semantic hash function that maps semantically
similar vectors to nearby intervals in an ordered key space while assigning denser regions
more key-space capacity. \emph{\VortexOverlayName} is a decentralized routing and state-management layer
that maintains reachability, ownership, and region metadata over that key space under weak
coordination. Each peer also runs a local ANN index engine over the vectors it currently serves.

\smallskip
\noindent\textbf{Query path.}
A query may arrive at any peer. The ingress peer evaluates \VortexHashName, routes to
the owner of the resulting key through the \VortexOverlayName, expands around that anchor with bounded
regional expansion, and queries the
resulting shortlist of peers to execute peer-local ANN search. The ingress peer then
merges the returned candidates and retrieves the full ANN query result. 

\smallskip
\noindent\textbf{Control and maintenance path.}
New vectors are hashed under the active \VortexHashName{} epoch and routed to their current owners, which update local ANN indexes and refresh the affected region metadata. Peers
periodically exchange heartbeats that carry both overlay liveness information and compact
semantic synopses. Any live peer with sufficient synopsis state may construct and propose a new \VortexHashName{} epoch. Ordinary joins, leaves, and churn are handled by \VortexOverlayName{} repair and metadata maintenance, while epoch refresh is hysteretic, validated before activation, and kept off the query path.

\subsection{\VortexHashName: Learned Semantic Hashing}
\label{subsec:vortexhash}

\begin{figure}[htbp]
    \centering
    \VortexTighten{-2ex}
    \includegraphics[width=\linewidth]{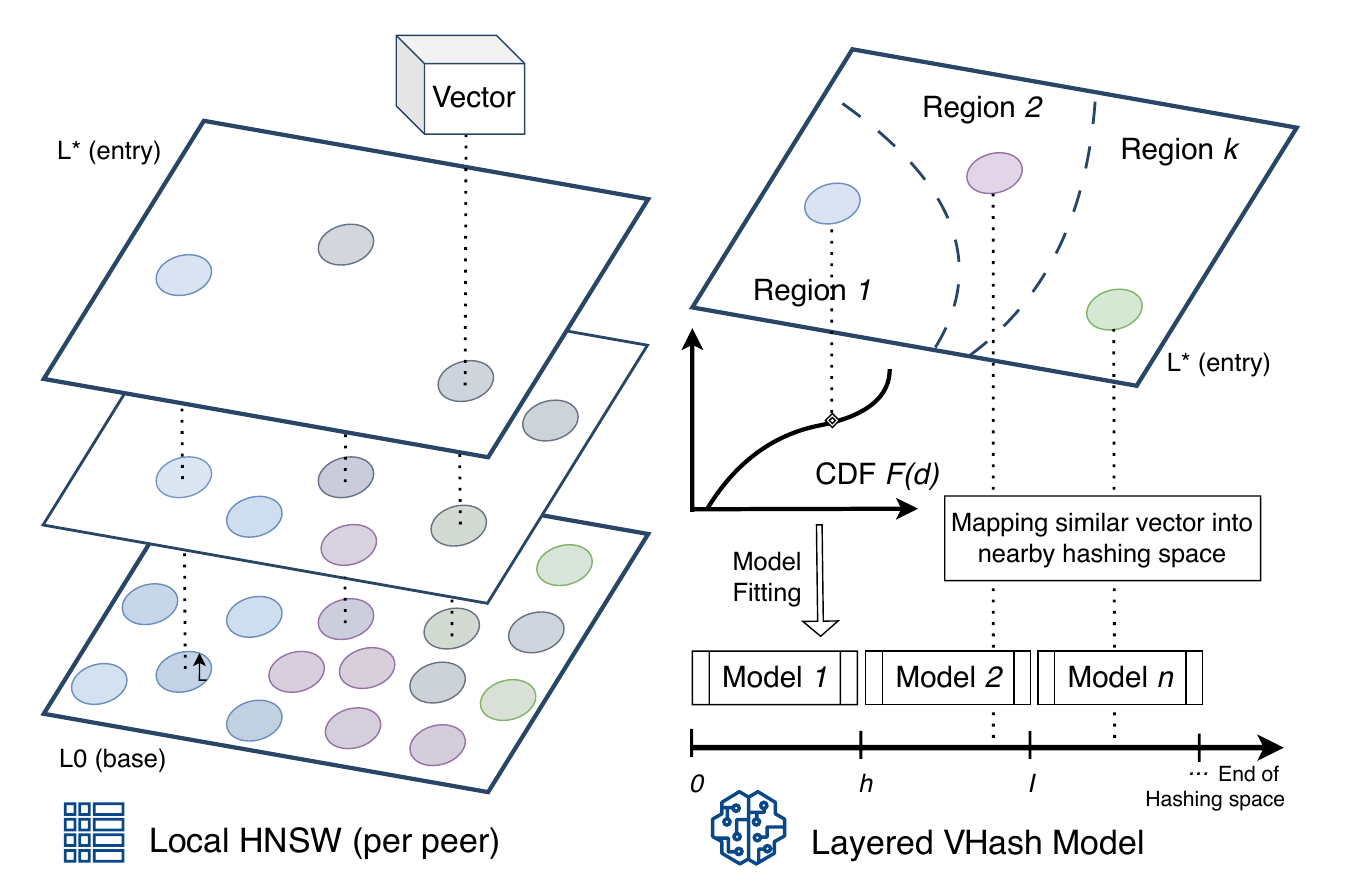}
    \Description{\VortexHashName{} pipeline showing coarse semantic partitioning, topology-aware region ordering, capacity allocation by region density, and learned in-region placement of vectors into an ordered overlay key space.}
    \VortexTighten{-4ex}
    \caption{\VortexHashName{} Design.}
    \VortexTighten{-3ex}
    \label{fig:vortexhash}
\end{figure}

Under epoch $e$, \VortexHashName{} instantiates a learned hash function
\begin{equation}
h_{\theta}^{(e)} : \mathbb{R}^{d} \rightarrow [0,2^{b}),
\end{equation}
whose output is the ordered overlay key used for placement and lookup. \VortexHashName's job is
to embed enough semantic structure into the key space that the overlay can narrow search
to a small candidate frontier. 

\VortexHashName{} does not encode the full nearest-neighbor relation. Instead, it constructs a semantics-preserving key space in which the owners of true neighbors tend to appear early enough in the semantic order of the vector space. The contract of \VortexHashName{} is intentionally weaker
than exact ANN and stronger than random hashing. It does not attempt to predict the final
top-$k$ neighbors directly. Instead, it satisfies two properties that matter for
decentralized retrieval. First, semantically nearby vectors should map to nearby hash
intervals often enough that ring neighborhoods are informative for search. Second, dense
semantic regions should receive proportionally more key-space capacity so that skew is
reduced before the overlay assigns ownership to physical peers. \VortexHashName's resulting semantic key order is a routing signal. The point is to make the key space informative enough that bounded correction to recover high recall without network-wide fanout.

\subsubsection{\VortexHashName{} Training and Refresh}
\label{subsec:vortexhash-construction}

\VortexHashName{} is trained as \emph{layered learned models}. It separates three construction tasks: learning coarse semantic regions, ordering and allocating intervals to those regions, and fitting an in-region coordinate. Training combines these components into the epoch-specific state
\begin{equation}
\theta^{(e)} =
\Big(
C,\pi,
\{(s_j,r_j,\widehat{f}^{(1)}_j,\{\widehat{f}^{(2)}_{j,\ell}\}_{\ell=0}^{B_j-1})\}_{j=1}^{K},
A
\Big),
\end{equation}
Here $C$ contains the $K$ centroids, $\pi$ is the learned region order, $(s_j,r_j)$ specifies a region's interval, $\widehat{f}^{(1)}_j$ is the first-level recursive model for region $j$, and $A$ identifies active regions. 

Training solves three different problems simultaneously: it must identify a coarse
semantic region, place that region into an ordered key space, and learn a fine-grained
coordinate within the region's interval. Formally, training complies the learned hash function
$h_{\theta}^{(e)} : \mathbb{R}^{d} \rightarrow [0,2^b)$.

Training is decentralized. Each peer maintains a local HNSW index over the vectors it
currently serves. On the query path, this index answers peer-local ANN queries. On the
control path, its upper layers already summarize the navigational structure
most relevant to overlay-scale search. \VortexSystemName{} reuses that structure. Peers periodically piggyback compact HNSW synopses on
the same heartbeat stream used for overlay maintenance. These synopses include
representative vectors together with navigational links and form a distributed compact summary of the scope's semantic layout. The synopsis budget is capped per peer. Any live peer may synthesize a new epoch from the synopses it has received. This role is transitive rather than a permanent coordinator.  What the data path requires is only that peers eventually activate the same compact hash state for the new epoch.

\smallskip
\noindent\textbf{Layer 1: coarse semantic partition.}
Let $\mathcal{S}_{\mathrm{syn}} \subset \mathbb{R}^{d}$ denote the current set of synopsis
vectors assembled for a refresh round. \VortexHashName{} first learns a coarse partition of
$\mathcal{S}_{\mathrm{syn}}$ into $K$ semantic regions with centroids
$C = \{c_1,\ldots,c_K\}$. Each synopsis vector $x \in \mathcal{S}_{\mathrm{syn}}$ is
assigned to its nearest centroid:
\begin{equation}
a(x) = \arg\min_{j \in [K]} \lVert x - c_j \rVert_2^2.
\end{equation}
These assignments define the represented mass of region $j$,
\begin{equation}
m_j = |\{x \in \mathcal{S}_{\mathrm{syn}} \mid a(x)=j\}|,
\end{equation}
and the set of centroid-local distances
\begin{equation}
D_j = \{\lVert x - c_j \rVert_2^2 \mid x \in \mathcal{S}_{\mathrm{syn}},\; a(x)=j\}.
\end{equation}
The purpose of this layer is to expose
coarse semantic regions that are stable enough to support routing.

\smallskip
\noindent\textbf{Layer 2: topology-aware order and capacity assignment.}
The second layer turns the learned regions into a routable one-dimensional key space.
\VortexHashName{} does not force a single global projection over a geometry that is rarely
globally linear. Instead, it induces a total order from topology. It builds a centroid
$k$-nearest-neighbor graph under exact squared Euclidean distance. \VortexSystemName{} then extracts a minimum spanning tree, performs a deterministic
depth-first traversal, removes obvious long jumps with a bounded 2-opt pass, and cuts the longest edge to obtain the linear region order
\[
\pi(1), \pi(2), \ldots, \pi(K),
\]
where $\pi(t)$ is the region identifier at position $t$ in the hash order.

This ordered region model is then complied into interval assignments in the $b$-bit key
space. Each active region $j$ receives a contiguous interval
$I_j = [s_j, s_j + r_j) \subset [0,2^b)$ whose width is proportional to represented mass:
\VortexTighten{-2ex}
\begin{equation}
r_j \approx 2^b \cdot \frac{m_j}{\sum_{u=1}^{K} m_u}.
\label{eq:range-allocation}
\end{equation}
Widths are rounded with a largest-remainder rule subject to a minimum width for active
regions. Regions with zero represented mass are marked inactive and excluded from
inference. The configured number of active regions is constrained so that the minimum-width
allocation fits in the $b$-bit key space, and the final interval is adjusted so that
active intervals form a disjoint partition of $[0,2^b)$. This layer is the stage of load balance in \VortexHashName. Dense semantic regions
receive more key-space capacity. 

\smallskip
\noindent\textbf{Layer 3: recursive in-region hash modeling.}
The first two layers determine \emph{which} semantic region a vector belongs to and
\emph{where} that region lies in the ordered key space. The third layer learns the
remaining degree: the vector's final \emph{hash rank within that region's interval}. 

For each region $j$, \VortexHashName{} models the distribution of centroid-relative distances
through the empirical radial CDF
\begin{equation}
F_j(d) = \Pr\!\left[\lVert x - c_j \rVert_2^2 \le d \mid a(x)=j\right].
\end{equation}
Once the region has been fixed, it becomes a structured one-dimensional
prediction problem: given
$d = \lVert x - c_j \rVert_2^2$, predict the percentile of the vector within region $j$'s
radial distribution. That percentile is the rank needed to place the vector
inside the interval already assigned to region $j$.

\VortexHashName{} therefore fits $F_j$ with a \emph{two-level recursive model}. Let $\widehat{f}^{(1)}_j : [0,\infty) \rightarrow [0,1)$ denote the
first-level model for region $j$, and let
$\{\widehat{f}^{(2)}_{j,\ell}\}_{\ell=0}^{B_j-1}$ denote the second-level models associated
with that region. The first-level model predicts a coarse percentile and routes the input
to one of $B_j$ specialized second-level models.
At inference time, percentile predictions are clamped to $[0,1-\epsilon]$ before interval materialization. 

The first-level model captures coarse radial position within region $j$ and routes the input to a local "expert" model. Each second-level model is then trained only on a restricted subdistribution of $D_j$, allowing it to fit the local shape of the region with lower error and smaller model cost than a single global  predictor. The result is a learned hash that is \emph{global where it must be} and \emph{local where it can be}: the upper layers learn the semantic and routable skeleton of the key space, while the recursive local models learn a lightweight ordering signal.

This is also where \VortexHashName{} differs from a conventional learned index.
Learned indexes operate over keys that are already ordered and use staged models to
predict positions in that pre-existing order. \VortexHashName{} uses the similar staged-model
principle for a different systems purpose. The semantic partition and topology-aware
interval assignment create the order; the recursive model then learns the residual
one-dimensional structure inside each learned region. 
It completes the learned hash by
converting centroid-relative structure into a stable overlay-space routing signal.

The decomposition has three advantages. First, it simplifies the learning
target: each second-level model solves a local monotone prediction problem instead of a
global vector-to-key regression. Second, it improves robustness under drift: changes in
the local density of one region can often be absorbed by retraining that region's
recursive model without perturbing the global region order. Third, it keeps the online
path small enough for direct use before overlay routing: inference requires region
selection, one first-level model evaluation, one selected second-level model evaluation,
and interval scaling.

\smallskip
\noindent\textbf{Epoch output and refresh.}
The output of training is a compact, versioned \VortexHashName{} epoch states. This compact state
is disseminated through the overlay as the next hash epoch. Peers activate a new epoch
only after receiving the complete compact state
and validating its digest, so all routing decisions within an epoch are evaluated against the same \VortexHashName. \VortexSystemName{} assumes authenticated control messages. Epoch identifiers are monotonically increasing within a scope. If multiple candidate epochs are proposed, peers accept the highest valid epoch under a deterministic tie-breaker on the epoch digest. This rule provides eventual convergence among live peers lookup messages carry an epoch number, and receivers reject requests for epochs that are neither active nor draining. Queries may
consult active and draining epochs during rollout; writes use the active write epoch. \VortexAppendixRefCap{app:vortexhash-selection} describes learned hash model selection and scoring.

\subsubsection{Hash Inference}
\label{subsec:vortexhash-inference}

Inference follows the same layered path as training.

\smallskip
\noindent\textbf{Step 1: region selection.}
Given a query or insert vector $v$, the peer first selects the nearest active region:
\begin{equation}
j^*(v) = \arg\min_{j \in A} \lVert v - c_j \rVert_2^2.
\end{equation}
Let
\VortexTighten{-4ex}
\begin{equation}
d^*(v) = \lVert v - c_{j^*(v)} \rVert_2^2
\end{equation}
denote the centroid-relative distance to the selected region. This step determines the
coarse semantic region in which $v$ will be placed.

\smallskip
\noindent\textbf{Step 2: recursive in-region model evaluation.}
Conditioned on the selected region, \VortexHashName{} evaluates the learned recursive model for
that region. The first-level model predicts a coarse in-region percentile and routes
the input to one of $B_{j^*(v)}$ specialized second-level models. The selected second-level model then outputs the final in-region percentile. \VortexHashName{} materializes the clipped percentile, which is interpreted as the learned coordinate of $v$ within region
$j^*(v)$'s interval. 

\smallskip
\noindent\textbf{Step 3: interval materialization.}
If region $j^*(v)$ owns interval start $s_{j^*(v)}$ and width $r_{j^*(v)}$, the final
\VortexHashName{} value is
\begin{equation}
h_{\theta}^{(e)}(v) =
s_{j^*(v)} +
\left\lfloor
r_{j^*(v)} \cdot \widetilde{F}_{j^*(v)}(d^*(v))
\right\rfloor.
\label{eq:vortex-hash}
\end{equation}

\smallskip
\noindent\textbf{Inference cost.}
The inference cost of \VortexHashName{} is small. Each peer stores only the compact
epoch state: the centroids and their order, per-region interval metadata, and the
parameters of the recursive in-region models. The footprint therefore scales with model
size rather than with the number of vectors stored in the system. Evaluating the hash
requires one nearest-region selection, one first-level recursive model evaluation, one
selected second-level model evaluation, and interval scaling with integer arithmetic. This
keeps hash inference cheap enough to execute directly on the latency-critical path before
any overlay routing or peer-local ANN work begins.


\subsection{\VortexOverlayName: Decentralized Routing and Live State over the Learned Key Space}
\label{subsec:vortex-vdht}

\VortexHashName{} decides \emph{where} semantic neighborhoods lie, and \VortexOverlayName{} decides \emph{how}
peers find and maintain them under weak coordination. \VortexOverlayName{} is a structured overlay over
the ordered \VortexHashName{} space. In a conventional DHT, key order supports ownership and
bounded exact-key lookup but carries no application semantics. In \VortexSystemName, nearby key
ranges correspond to nearby learned regions in the vector space. Hash value locality is
therefore useful both for locating owners and for restricting similarity search to a
small candidate frontier.

\VortexOverlayName{} stores two classes of decentralized state. First, \emph{vector ownership}: a vector $x$ is served
by the successor of $h_{\theta}^{(e)}(x)$ in the scope-qualified ring for epoch $e$.
Second, \emph{region-directory state}: for each active region $j$, \VortexSystemName{} maintains a
region peer set $P_j^{(e)}$ records the peers currently serving vectors from that region
under epoch $e$. 

When membership changes alter
which peers serve region $j$, the affected peers probe and refresh the corresponding directory
entry. No peer keeps a fully current map from all regions to all owners. Every
ingress peer computes immediate hash-based routing hints locally and resolves
slower-changing directory state through the overlay only.

\subsubsection{Bounded-fanout ANN Query Processing}
\label{subsec:vortex-query}

\VortexSystemName{} resolves the selectivity-versus-robustness tension by consulting routing signals in
strictly increasing order of cost. The learned key is used first, ring-local repair is
used next, and region-directory expansion is invoked only when the cheaper signals are
insufficient.

Let $o(q)$ denote the successor owner of the query key $h_{\theta}^{(e)}(q)$ under the
current epoch. Let $N_{\rho}(o(q))$ denote the bounded predecessor/successor neighborhood
of radius $\rho$ around that owner, and let $R_r(q)$ denote the top $r$ region hints
returned by centroid distance under $\mathcal{H}^{(e)}$. The logical candidate frontier is
\begin{equation}
\mathcal{C}(q) =
\operatorname{Budget}_{B}
\left(
\{o(q)\}
\;\cup\;
N_{\rho}(o(q))
\;\cup\;
\bigcup_{j \in R_r(q)} P_j^{(e)}
\right),
\label{eq:candidate-frontier}
\end{equation}
where $B = \texttt{max\_query\_nodes}$ and $\operatorname{Budget}_{B}$ retains at most $B$
distinct peers after deduplication in stage order. In the implementation, these probes are
staged and may stop early; Equation~\ref{eq:candidate-frontier} captures the design
contract. When a region peer set exceeds the remaining budget, \VortexSystemName{} orders entries by
regional rank, interval overlap with the query key, liveness, and recent load, and
then admits peers until the global budget is reached. 

A query may enter at any peer. The ingress peer first evaluates \VortexHashName{} to obtain
$h_{\theta}^{(e)}(q)$ and a short ranking of nearby regions. It then resolves $o(q)$
through the \VortexOverlayName. This peer is the initial anchor. Because \VortexHashName{} preserves coarse
semantic locality, the anchor is often already close to the correct serving set.

\VortexSystemName{} treats learned placement as predictive rather than exact. Queries near
region boundaries, or queries whose true neighbors span multiple adjacent regions, may not
be well served by the anchor alone. \VortexSystemName{} therefore next probes a bounded
predecessor/successor neighborhood around the anchor. This ring-local repair stage is
cheap and corrects the common case in which the right peer lies just across a
learned-region or ownership boundary.

If more evidence is needed, the ingress peer performs region-directory expansion. It
resolves the region peer sets associated with the top region hints and merges the returned
peers into the candidate frontier. This is the main semantic correction mechanism in the
system. Importantly, it remains fully decentralized: the directory itself lives in the
\VortexOverlayName{} and is reached through ordinary overlay lookup rather than through a separate
planner, metadata service, or query router.

\VortexSystemName{} does not force every query to pay the same routing cost. The ingress peer estimates
query ambiguity from the gap between the two best region distances:
\begin{equation}
\Delta(q) =
\lVert q - c_{(2)} \rVert_2^2 -
\lVert q - c_{(1)} \rVert_2^2,
\end{equation}
where $c_{(1)}$ and $c_{(2)}$ are the nearest and second-nearest active region centroids. Large gaps indicate clear region membership and justify smaller $\rho$ and $r$. Small gaps indicate boundary ambiguity and trigger broader ring-local expansion and more directory probes. Routing
effort is therefore elastic within the hard fanout cap.

The selected peers then execute peer-local ANN search over the vectors they
currently serve and return scored candidates together with payload handles or references.
The ingress peer merges these partial results into the final top-$k$ results. 

\subsubsection{Churn Resistance and Maintenance}
\label{subsec:vortex-maintenance}

This section states the query-path robustness and liveness properties of
\VortexSystemName{} under network churns. \VortexAppendixRefCap{app:maintenance-details} gives the detailed designs for routing information maintenance and caching.

\smallskip
\noindent\textbf{Robustness under churn.}
\VortexSystemName{} inherits the standard structured-overlay robustness. Successor
pointers and successor lists determine eventual reachability of the owner for a key range. Region-directory records are resolved through the same live overlay and cached only as an optimization. As a result, stale routing results may lengthen routes and stale caches may force a live metadata lookup, but neither changes the authoritative owner or region record. During churn, lookups continue against the current ring structure maintained by stabilization and successor repair. During epoch rollover, active and draining state are kept disjoint by epoch tags. Under eventual heartbeat delivery among live peers, the overlay converges back to a consistent status, and query processing continues to make progress throughout churn and rollout.

As the system property, the common path remains fast because hot routing state can be served from local caches, while correctness and recovery depend only on the liveness of \VortexOverlayName. \VortexSystemName{} therefore does not require a centralized metadata service, or a controller. As long as the overlay can repair ownership and metadata resolution under failures, query processing remains available.

\section{Evaluation}
\label{sec:evaluation}

We evaluate \VortexSystemName{} as an end-to-end distributed retrieval system, not just as an
isolated partitioner or ANN primitive. The evaluation asks four questions.

First, does \VortexSystemName{} improve recall for a fixed budget of contacted peers,
hot-path messages, and overlay hops?

Second, at the same recall, does this
selectivity reduce end-to-end latency, communication, and retrieval-stage
fanout?

Third, does \VortexSystemName{} remain live and efficient under skewed data, hot
queries, churn, failures, and \VortexHashName{} epoch updates?

Fourth, as the system
and workloads grow, does \VortexSystemName{} preserve recall and latency while keeping
selection metadata, peer-local ANN memory, and control-plane overhead bounded?

Together, these experiments test whether \VortexSystemName{} turns routable semantic
locality into systems benefits: high recall with bounded fanout, lower
retrieval cost, balanced load under skew, lower peer-local ANN memory
requirement, and liveness under churn and refresh. \VortexSystemName{} targets weakly
coordinated deployments where queries may enter at arbitrary peers and no
always-current trusted router is available. Rather than trying to dominate a
centralized vector system with an acceptable trusted router, we ask how much of
the benefit of semantic routing \VortexSystemName{} can preserve while keeping ownership,
peer discovery, repair, and metadata maintenance decentralized.

\subsection{Methodology}
\label{subsec:eval-methodology}

\noindent\textbf{Hardware and environments.}
We implemented \VortexSystemName~\cite{VortexSystemRepo,VortexHashRepo} and all baselines both as networked applications and in an
event-driven evaluation substrate built on OverlaySim~\cite{OverlaySimRepo}. The goal is to realize each method as a live distributed
control plane with explicit query injection, message delivery, timer firing,
topology-aware delay, transport overhead, and failure or churn events. Each
experiment is compiled into an executable event graph and run by the same
simulation kernel. Queries are issued as time-stamped ingress events, method
logic runs inside service modules, and control-path state evolves through
messages and timers rather than through offline replay. Unless otherwise
stated, controlled simulations on 1M--10M-vector workloads ran on an 32-vCPU
KVM with 3.0\,GHz virtual cores and 120\,GB of DDR5 system memory.

We also validate \VortexSystemName{} on a 32-peer public cloud testbed: eight servers in the California, four logical peers per server, and 19\,ms mean inter-machine RTT. The same service modules run as networked applications over real RPC, serialization, scheduling, and transport, testing whether the controlled-evaluation trends hold in deployment.

\smallskip
\noindent\textbf{Baselines.}
We compare against four baselines spanning the main alternatives for weakly
coordinated retrieval: exact-key decentralized routing, decentralized
LSH-style bucketing, hierarchical partition routing, and coordinator-aided
semantic routing. All methods use the same peer-local ANN backend, timeout
policy, merge logic, and retrieval model.

\textsc{RandomSelector} is the no-semantic-placement overlay baseline. Base
vectors are placed by consistent hashing; queries receive the same
candidate-peer budget as \VortexSystemName{} and form a capped frontier through bounded
successor or ring-neighbor exploration. This baseline tests whether a
decentralized exact-key overlay with fair fanout is already sufficient.

\textsc{Centralized Router} is a coordinator-aided upper-bound reference with
fresh centralized routing metadata over coarse routing units. It dispatches
queries to selected workers before the common merge stage. We include it to quantify what a fresh
trusted routing tier can buy, and to measure how closely \VortexSystemName{} approaches that
frontier while preserving decentralized ownership and lookup.

\textsc{Distributed ScaNN} represents a hierarchical partition-routing design
derived from ScaNN's partition-and-score architecture~\cite{Guo2020ScaNN}. Our
implementation externalizes ScaNN's coarse routing stage into explicit router
nodes and leaf partitions, thereby evaluating the distributed systems design
point suggested by hierarchical partitioned search.

\textsc{DET-LSH} adapts DET-LSH-style semantic buckets to the same overlay,
following DET-LSH~\cite{Wei2024DETLSH}. Each query generates LSH keys, resolves
bucket owners through \VortexOverlayName, retrieves bucket metadata, ranks owner candidates,
and dispatches a capped set of shard searches. This baseline tests whether
decentralized semantic grouping via traditional LSH is sufficient without a
stronger routable learned hash function.

Finally, \textsc{\VortexSystemName} combines \VortexHashName{} with the \VortexOverlayName{} overlay described in
Section~\ref{sec:design}.
Taken together, these baselines cover the comparison space that matters for this
application scenario. \textsc{RandomSelector} tests exact-key decentralization
without semantic placement. \textsc{Centralized Router} tests
coordination-favored semantic routing. \textsc{Distributed ScaNN} tests
hierarchical partition routing. \textsc{DET-LSH} tests decentralized semantic
bucketing with traditional LSH.

\smallskip
\noindent\textbf{Datasets.}
We evaluate \VortexSystemName{} on six real-world vector workloads spanning standard ANN benchmarks, high-dimensional descriptors, neural embeddings, production-style web-search embeddings, and a billion-scale endpoint in \VortexDatasetAppendixRef{}. SIFT1M and GIST1M provide standard million-scale workloads from the ANN benchmarking literature
\cite{Aumuller2020ANNBenchmarks,TexMexCorpus,Jegou2011ProductQuantization}.
SIFT10M is drawn from the BigANN/SIFT benchmark family
\cite{TexMexCorpus,Simhadri2022BigANN}. DEEP10M-L2 is a 10M-vector slice of the
Deep1B benchmark evaluated under its Euclidean-distance convention
\cite{Babenko2016Deep1B,Simhadri2022BigANN}. SPACEV10M and SPACEV1B are drawn
from Microsoft's SPACEV1B web-search vector dataset, whose released benchmark
definition includes document vectors, query vectors, historical queries, and
L2 nearest-neighbor ground truth~\cite{MicrosoftSPACEV1B,Simhadri2022BigANN}.

\smallskip
\noindent\textbf{Latency model.}
We use the NetLatency-Data PlanetLab RTT matrices as the empirical
network-latency model~\cite{Zhu2017NetLatency,NetLatencyData}. The traces are
used only to instantiate link delay in OverlaySim. For the main experiments,
we use a Regional effective-latency regime by uniformly scaling the measured
one-way delays by 0.05. This preserves the relative heterogeneity of the
measured topology while modeling regional deployments where routing and fanout
costs matter but are not overwhelmed by global-WAN delay. Hot-path messages
count query-caused routing, directory lookup, peer dispatch, and merge messages;
they exclude periodic maintenance, model dissemination, and epoch-refresh
traffic.

\VortexAppendixRefCap{app:workloads} summarizes workload sizes, ground-truth accounting,
and latency-trace handling.

\VortexTighten{-1ex}
\subsection{Testbed Performance}
\label{subsec:eval-testbed}

We begin with the physical deployment. The testbed runs the same \VortexSystemName{} and
baseline service modules used in OverlaySim~\cite{OverlaySimRepo} as networked applications. This
experiment checks whether the main recall, latency, memory, and scoped-retrieval
trends persist under real RPC, serialization, host scheduling, and transport
effects.

\begin{figure*}[!t]
  \centering
  \includegraphics[width=\linewidth]{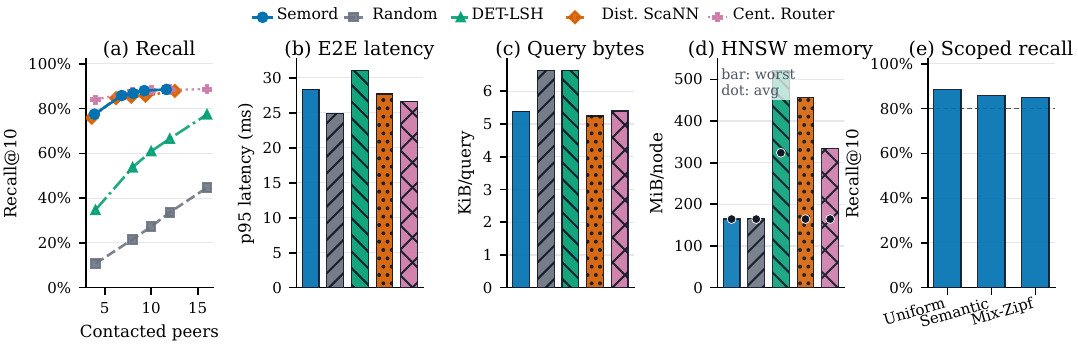}
  \Description{Testbed performance figure showing recall, latency, load balance,
  and robustness measurements for the physical deployment.}
  \VortexTighten{-4ex}
  \caption{\textbf{Testbed performance.}
  \VortexSystemName{} preserves high recall with low latency, balanced load, and robustness in the deployment.}
  \label{fig:testbed}
\end{figure*}

Figure~\ref{fig:testbed} validates the headline behavior in deployment.
Panel~(a) shows the recall--fanout frontier. \VortexSystemName{} reaches the high-recall
region with a small contacted-peer frontier. This confirms that the learned semantic
frontier remains useful after replacing simulated message delivery with real networked processes. Panel~(b) and (c) reports p95 end-to-end latency and query bytes. \textsc{Centralized Router} is the lowest-latency reference because it uses fresh centralized routing state. \VortexTestbedLatencySubject{} closely matches the best centralized baseline latency at this scale. \VortexSystemName{} is competitive with the high-recall distributed methods while preserving decentralized ownership and lookup. Panel~(d) reports peer-local HNSW memory. \VortexSystemName{} keeps both average and worst-node HNSW memory low and balanced, comparable to other baselines. Panel~(e) reports the scoped-retrieval recall on the physical deployment.
Across Uniform, Semantic, and Mixed-Zipf scopes, \VortexSystemName{} remains above the scoped-recall target. The remaining experiments use the controlled environment to sweep larger scale, skew, churn, and epoch-refresh regimes beyond the physical testbed.  \VortexSystemName{} improves recall by more than 15\% and reduces contacted peers by over 60\% compared with decentralized baselines. \VortexSystemName{} also approaches the recall and latency of a centralized oracle baseline while reducing peak peer-local ANN index memory by more than 2$\times$.
\subsection{Selectivity and Matched-Recall Cost}
\label{subsec:eval-selectivity-cost}

The first two questions are tightly coupled: \VortexSystemName{} should improve recall per
routing budget, and that selectivity should reduce end-to-end cost at the same
retrieval quality. We sweep the routing budget $B$ and measure recall@10,
realized peer fanout, and hot-path messages. We then compare end-to-end latency
at a fixed recall target.

\begin{figure*}[!t]
\centering
\includegraphics[width=\linewidth]{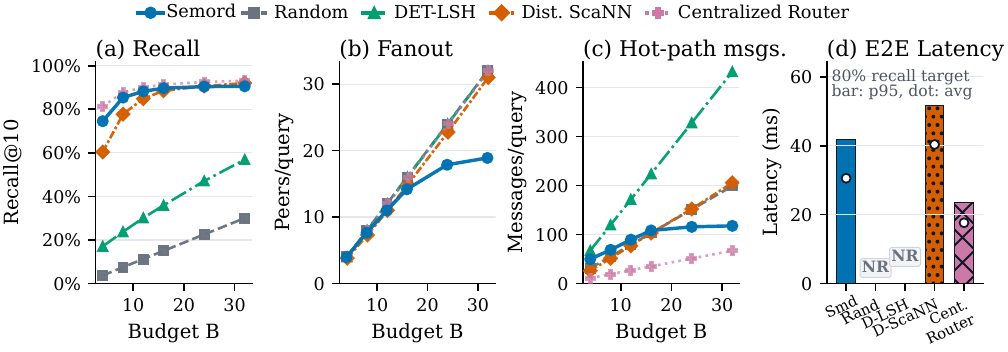}
\Description{Multi-panel recall-cost frontier comparing methods by recall at increasing routing budgets, realized peer fanout, hot-path messages, and matched-recall latency.}
\VortexTighten{-4ex}
\caption{\textbf{Recall-cost frontier.}
\VortexSystemName{} reaches high recall with low fanout and few hot-path messages; NR means not reached.}
\VortexTighten{-2ex}
\label{fig:selectivity-cost}
\end{figure*}

Figure~\ref{fig:selectivity-cost} is the main recall-cost result. \VortexSystemName{} reaches
the high recall with a smaller effective frontier than the decentralized
baselines. \textsc{RandomSelector} spends budget on semantically arbitrary peers
and does not reach the 80\% recall target. \textsc{DET-LSH} improves over random
placement, but its unordered buckets require many hot-path messages and still
do not reach the target. \textsc{Distributed ScaNN} reaches high recall, but
only with near-full fanout and substantially higher message cost. \VortexSystemName{} closely tracks \textsc{Centralized Router} in recall while preserving
decentralized routing. Its realized fanout grows slowly with $B$ because
\VortexHashName{} exposes a compact semantic frontier and \VortexOverlayName{} expands only as needed.
This is the payoff of making semantic locality routable: the overlay spends its
budget on useful peers rather than broad fanout. Panel~(d) compares methods at the 80\% recall target. \textsc{Centralized
Router} is fastest, as expected, because it assumes fresh centralized routing
state and avoids decentralized owner discovery. \textsc{Distributed ScaNN}
reaches the target but pays higher fanout and message cost, which leads to
higher tail latency than \VortexSystemName{} in this setting. \VortexSystemName{} remains close while
preserving decentralized ownership, lookup, and repair through \VortexOverlayName. Thus, the
gap quantifies the cost of decentralization, while the gap to
deployable distributed baselines shows the benefit of \VortexSystemName's learned semantic
overlay.

\subsection{Scoped Retrieval}
\label{subsec:eval-scoped}

\VortexSystemName{} targets scoped retrieval, where the scope is both an eligibility
constraint and a routing boundary. We therefore evaluate \VortexSystemName{} with exact
ground truth computed over the scope-eligible vector set. This experiment checks whether the learned overlay remains effective when scope construction varies from
semantically coherent to weakly aligned with embedding locality.
\begin{figure}[htbp]
  \centering
  \includegraphics[width=\linewidth]{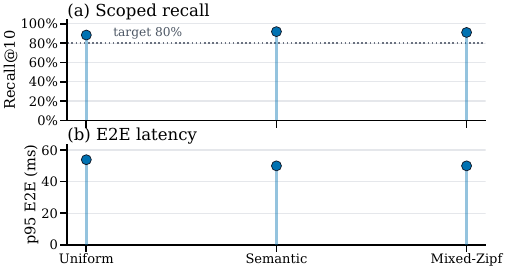}
  \Description{Scoped retrieval figure showing \VortexSystemName{} recall@10 and p95
  end-to-end latency across Uniform, Semantic, and Mixed-Zipf scope regimes.}
  \VortexTighten{-3ex}
  \caption{\textbf{Scoped retrieval.}}
  \VortexTighten{-6ex}
  \label{fig:scoped-retrieval}
\end{figure}
Figure~\ref{fig:scoped-retrieval} shows that \VortexSystemName{} remains effective across
all three scope constructions at $B{=}16$. Panel~(a) reports scoped recall@10:
Uniform, Semantic, and Mixed-Zipf scopes all exceed the 80\% target. The Uniform
case is the most conservative regime because scope membership is weakly aligned
with embedding locality; \VortexSystemName{} still reaches high recall without relying on
scopes that are already semantically clustered. Semantic and Mixed-Zipf scopes
represent the intended operating regimes for workspace-like collections with
topical structure and outliers. Panel~(b) shows that enforcing scope does not destabilize the serving path. Across the three regimes, p95 end-to-end latency remains nearly flat.

\subsection{Why Routing Becomes Selective}
\label{subsec:eval-why}

The selectivity result depends on whether \VortexHashName{} produces a useful overlay
key. A learned key $h_\theta(q)$ is useful only if it routes the query toward
the peers that store the exact nearest neighbors. We test this directly: for
each query, we compute GT@10, identify the true serving peers, and measure
whether the frontier reaches those responsible owners under a bounded budget.

\begin{figure}[htbp]
  \centering
  \includegraphics[width=\linewidth]{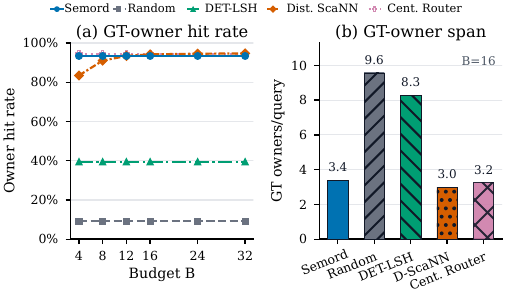}
  \Description{Two-panel \VortexHashName{} routing-signal figure showing responsible-owner hit rate and GT-owner span across routing methods.}
  \VortexTighten{-5ex}
  \caption{\textbf{\VortexHashName{} routing signal.}}
  \VortexTighten{-2ex}
  \label{fig:vortexhash-signal}
\end{figure}

Figure~\ref{fig:vortexhash-signal} measures whether \VortexHashName{} gives \VortexOverlayName{} a
useful routing signal. Panel~(a) reports responsible-owner hit: whether the
bounded frontier reaches a peer that owns a GT@10 vector. \VortexSystemName{} tracks
\textsc{Centralized Router} and \textsc{Distributed ScaNN} closely, while
\textsc{RandomSelector} and \textsc{DET-LSH} miss many responsible owners.
Owner hit is not the final retrieval metric; it explains why the full
recall-cost path in Figure~\ref{fig:selectivity-cost} improves after peer-local
ANN and merge.
Panel~(b) reports the GT-owner span at $B{=}16$: the number of distinct peers
holding the exact GT@10 answers. \VortexSystemName{} keeps this span small. This matters for decentralized retrieval because reaching the
right semantic region is not enough; the answer set must also be collectable
from a small number of owners. \VortexHashName{} provides both properties: it points to
the right region with \VortexOverlayName{} and keeps the true answer set concentrated enough
for bounded fanout.
\begin{figure}[htbp]
  \centering
\includegraphics[width=\linewidth]{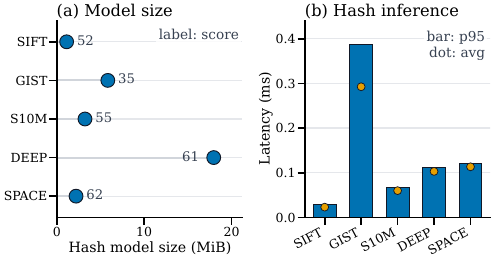}
  \Description{Two-panel \VortexHashName{} overhead figure showing selected model size, placement score, and per-query hash inference latency.}
  \VortexTighten{-4ex}
  \caption{\textbf{\VortexHashName{} quality and overhead.}}
  \label{fig:vortexhash-overhead}
\end{figure}
Figure~\ref{fig:vortexhash-overhead} shows that the learned hash is compact enough
for the hot path. The selected models remain compact across workloads, and
hash inference stays well below 1\,ms at both mean and p95. \VortexHashName{} therefore
adds a routing signal without turning the query path into model-serving work.

\subsection{Skew, Churn, and Refresh}
\label{subsec:eval-stress}

The third question is whether \VortexSystemName{} remains live and efficient under the
conditions that make weak coordination difficult. We evaluate three risks.
Skew tests whether selective routing creates serving hotspots. Churn tests
whether queries continue to make progress as peers leave or crash. Refresh
tests whether \VortexHashName{} can be updated as data grows without disrupting the
data path.
\begin{figure}[htbp]
  \centering
\includegraphics[width=\linewidth]{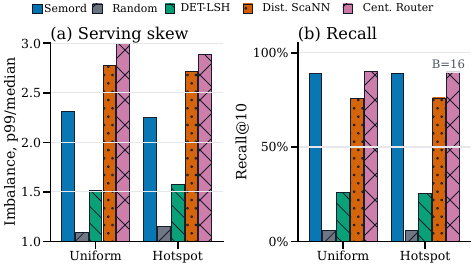}
  \Description{Two-panel skew figure showing serving-load imbalance and recall under uniform and hotspot query distributions.}
  \VortexTighten{-4ex}
  \caption{\textbf{Balance under skew.}}
  \label{fig:balance}
\end{figure}

Figure~\ref{fig:balance} shows that \VortexSystemName{} remains balanced under query skew.
Panel~(a) reports serving imbalance as p99/median load. Under both uniform and
hotspot query distributions, \VortexSystemName{} is substantially more balanced than
\textsc{Distributed ScaNN} and \textsc{Centralized Router}, while remaining
well below the extreme concentration that would indicate serving hotspots.
Panel~(b) shows that this balance is not achieved by giving up retrieval
quality: at $B{=}16$, \VortexSystemName{} maintains high recall under both workloads. This
distinction matters. A system can reduce imbalance only by spreading traffic
through broad fanout or by lowering recall. \VortexSystemName{} avoids both failure modes:
it preserves high recall while keeping hot-query amplification moderate.

\begin{figure}[htbp]
  \centering
\includegraphics[width=\linewidth]{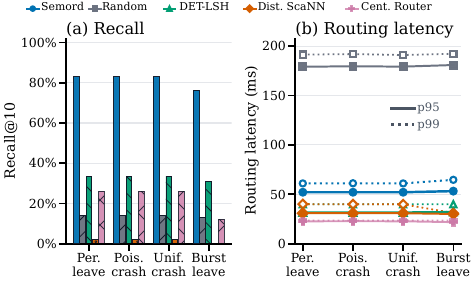}
  \Description{Churn-resilience figure comparing recall and routing latency across periodic leaves, Poisson crashes, uniform crashes, and bursty leave events.}
  \VortexTighten{-3ex}
  \caption{\textbf{Churn resilience.}}
  \label{fig:churn-failure}
\end{figure}

Figure~\ref{fig:churn-failure} shows that \VortexSystemName{} degrades gracefully under
membership change. We evaluate periodic leaves, Poisson crashes, uniform
crashes, and bursty leave events. Panel~(a) shows that \VortexSystemName{} maintains much
higher recall than the decentralized baselines under all four regimes, while
panel~(b) shows that its routing latency remains stable at both p95 and p99.
This isolates the effect of membership change on owner discovery and repair.
This behavior follows from \VortexSystemName's decentralized control plane. There is no
single router or globally current shard map whose failure makes queries
unroutable. Owner lookup and repair continue through the live \VortexOverlayName, so churn
degrades routing quality gradually rather than disabling peer discovery.
Failures may increase retries or force rerouting, but queries continue through
surviving overlay paths without falling back to network-wide scatter--gather.

\begin{figure}[htbp]
  \centering
  \includegraphics[width=\linewidth]{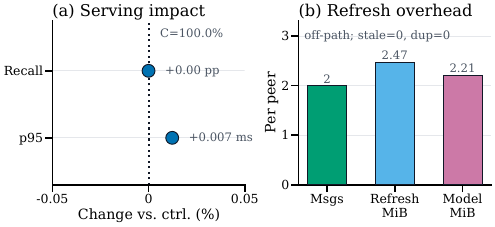}
  \Description{Hash-model refresh figure showing serving impact during refresh and background refresh cost in messages, transferred state, stale results, and duplicate candidates.}
  \VortexTighten{-4ex}
  \caption{\textbf{Hash-model refresh.}}
  \label{fig:epoch-refresh}
\end{figure}

Figure~\ref{fig:epoch-refresh} shows that \VortexHashName{} refresh is a live
maintenance operation rather than a stop-the-world rebuild. Panel~(a) measures
serving impact relative to a no-refresh control. Recall is unchanged, and p95
latency increases by only $0.007$\,ms, showing that refresh stays effectively
off path. Panel~(b) reports the refresh cost directly: a small message count and
light transferred state, with no stale results and no duplicate candidates
during the measured transition.

This is the key systems property: learned-state maintenance does not interrupt
queries or hide its cost inside the query path. Instead, refresh is exposed as
bounded background work while the serving path remains stable.

\subsection{Scale, Overhead, and Robustness}
\label{subsec:eval-scale-overhead}

The fourth question is whether \VortexSystemName{} remains practical as the system and
workloads grow. We measure scaling behavior, metadata and memory footprint, and
workload robustness. A peer group is the ownership and serving unit in the
controlled scale experiments; each group runs one logical peer-local ANN service
and participates in \VortexOverlayName{} routing.

\begin{figure*}[htbp]
  \centering
  \includegraphics[width=\linewidth]{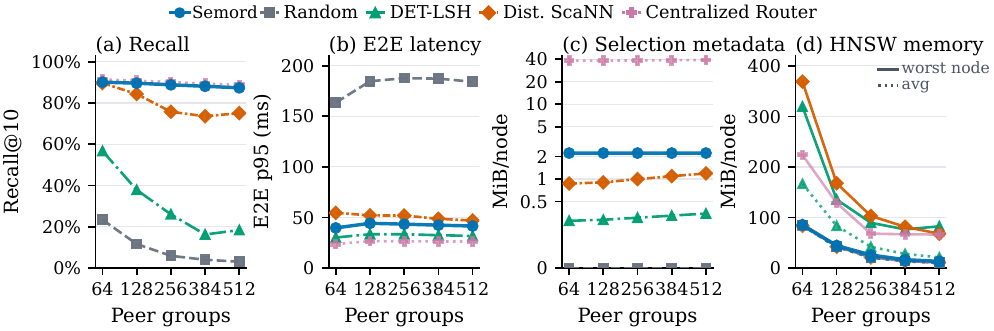}
  \Description{Scale and overhead figure showing recall, p95 latency, selection metadata, peer-local ANN memory, and control-plane overhead as the number of peer groups grows.}
  \VortexTighten{-3ex}
  \caption{\textbf{Scale and system overhead.}}
  \label{fig:scale-overhead}
\end{figure*}

Figure~\ref{fig:scale-overhead} shows that \VortexSystemName{} scales without relying on
hidden global state. As the number of peer groups increases, \VortexSystemName{} maintains
high recall and stable p95 latency with \VortexHashName{} and \VortexOverlayName.
\textsc{RandomSelector} has almost no selection metadata but fails to reach the
right peers. \textsc{DET-LSH} appears fast only because recall collapses.
\textsc{Distributed ScaNN} is competitive at small scale, but its recall drops
as the partition frontier becomes harder to cover.

The peer-local HNSW panel shows that \VortexSystemName's local ANN memory scales with
ownership, not corpus size. As the corpus is spread across more peer groups,
both average and worst-node memory fall, keeping per-peer indexes small and
balanced without replicating a global ANN structure.

\begin{figure}[htbp]
  \centering
  \includegraphics[width=\linewidth]{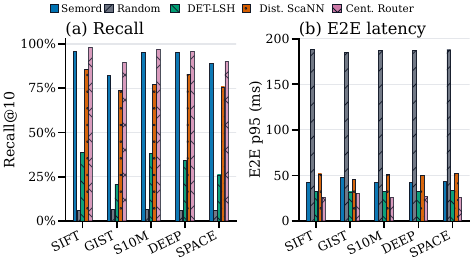}
  \Description{Workload robustness figure comparing recall and latency across SIFT, GIST, DEEP, and SPACEV workloads.}
  \VortexTighten{-3ex}
  \caption{\textbf{Workload robustness.}}
  \vspace{-2ex}
  \label{fig:workload-robustness}
\end{figure}

Figure~\ref{fig:workload-robustness} shows that \VortexSystemName's advantage is not tied
to one embedding distribution. Across SIFT, GIST, DEEP, and SPACEV workloads,
\VortexSystemName{} stays close to \textsc{Centralized Router} in recall while using a
decentralized query path. \textsc{RandomSelector} and \textsc{DET-LSH} are fast
only when they miss much of GT@10. \textsc{Distributed ScaNN} reaches moderate
to high recall, but \VortexSystemName{} provides the stronger recall-latency tradeoff across
workloads.

The appendix \ref{app:protocol-eval-details} reports additional evaluation results, including tail-latency and retrieval-cost breakdowns, SPACEV1B scale-up results on 1{,}024 nodes, and extended experimental details.

\section{Related Work}
\label{sec:related}

\noindent\textbf{Coordinated vector systems and semantic routing.}
Most production vector systems assume a coordinated routing layer that assigns
data, tracks shard state, dispatches queries, and merges partial results.
Milvus represents this design point for distributed vector data management,
VBase integrates vector retrieval with relational query processing, and Vexless
uses an orchestrator to assign serverless vector-search work
\cite{Wang2021Milvus,Zhang2023VBase,Su2024Vexless}. Coordinator-worker
similarity-search systems such as Pyramid make the same assumption: a trusted
routing tier has enough global structure to decide where to search
\cite{Deng2019Pyramid}. Our \textsc{Distributed ScaNN} baseline adapts ScaNN's
partition-and-score design as a hierarchical routing point
\cite{Guo2020ScaNN}. These systems are effective when fresh routing state is
available. \VortexSystemName{} targets a different boundary: queries may enter at arbitrary
peers, scope is itself a routing constraint, and the system must discover a
small promising peer set without a permanent trusted router or globally current
shard map.

\smallskip
\noindent\textbf{Structured and semantic overlays.}
Structured overlays such as Chord provide scalable decentralized exact-key
lookup, but randomized placement destroys the semantic locality needed for
similarity search \cite{Stoica2001Chord}. Ordered overlays such as SkipNet and
Mercury show that decentralized systems can preserve structure over names or
attributes while supporting scalable routing and load management
\cite{Harvey2003SkipNet,Bharambe2004Mercury}. Earlier semantic overlays such
as pSearch and Semantic Small World also recognized that content-aware
organization can reduce peer-to-peer search cost
\cite{Tang2003PSearch,Li2004SSW}. \VortexSystemName{} differs in the
target workload and routing contract: it learns an ordered key space from
dense-vector geometry, makes that key space executable by a live overlay, and
uses bounded repair plus peer-local ANN to recover high recall without
scatter--gather.

\smallskip
\noindent\textbf{Learned indexing, learned hashing, and locality-sensitive hashing.}
LSH and $p$-stable LSH provide probabilistic locality by mapping nearby vectors
to shared buckets \cite{Indyk1998LSH,Datar2004PStable}. This makes them natural
for decentralized bucket lookup, and our \textsc{DET-LSH} baseline tests this
design point \cite{Wei2024DETLSH}. Learned index structures use models to predict positions in an already ordered
key domain \cite{Kraska2018LearnedIndex}. Recent learned-DHT work such as LEAD
embeds learned models into structured overlays to accelerate distributed range
queries over ordered key-value data \cite{Wang2025LEAD}. \VortexHashName{} uses learned models for a different purpose: it learns the order from vector geometry and uses that order as a routing signal for decentralized vector retrieval.

\section{Conclusion}
\label{sec:conclusion}

\VortexSystemName{} realized a practical efficient decentralized vector overlay for similarity search and delivery. \VortexSystemName{} achieves high recall, preserves liveness under network churn, and keeps selection metadata and peer-local ANN memory bounded as the system grows. Its key idea is to make
semantic locality routable. \VortexHashName{} learns a compact, load-aware hash function that maps vectors into an ordered overlay key space; \VortexOverlayName{} then keeps ownership, metadata, and
epoch updates live and decentralized.

The broader lesson is that learned hash functions can make semantic structure directly useful for distributed placement, routing, and retrieval. 

The artifact is organized across three open-source projects.
\emph{LearnedHash}, part of the broader \emph{Learned Hash Function} library and ecosystem, contains the \VortexHashName{} implementation and model-selection modules~\cite{VortexHashRepo}. \emph{Vortex System} contains the networked retrieval system benchmarks, baseline implementations, experiment configurations, and analysis pipelines~\cite{VortexSystemRepo}. \emph{OverlaySim} contains the event-driven substrate used for the simulated testbed~\cite{OverlaySimRepo}.
\ifvortexincludeacknowledgments
\section*{Acknowledgment}

The authors were partially supported by NSF Grants 2322919, 2420632, 2426031, and 2426940.

\fi
\clearpage
\VortexPrintBibliography
\ifvortexincludeappendix
\appendix
\clearpage
\appendix
\makeatletter
\setlength{\@dblfpsep}{12pt}
\makeatother
\section{Protocol and Evaluation Details}
\label{app:protocol-eval-details}

This appendix provides additional design details and supplementary materials
for \VortexSystemName. Section~\ref{app:maintenance-details} describes the maintenance protocol. Section~\ref{app:vortexhash-selection}
describes \VortexHashName{} model selection. Section~\ref{app:workloads}
describes the experiments details. Section~\ref{app:evaluation}
describes the supplementary results.

\subsection{Maintenance Protocol Details}
\label{app:maintenance-details}

The main paper states the query-path robustness and liveness properties of \VortexSystemName. This section gives the maintenance procedures behind those properties.

\paragraph{Heartbeats.}
Peers periodically exchange heartbeats that carry both overlay liveness state
and compact semantic synopses. Heartbeats refresh successor/predecessor and
finger-table liveness, advertise region-activity changes, and piggyback the
upper-layer HNSW summaries used to maintain the distributed semantic synopsis.
The same control stream therefore keeps the overlay live and supplies the data
used to refresh the learned hash.

Epoch synthesis is gated by hysteresis on aggregate semantic or load drift
rather than by per-event triggers. This prevents churn-induced oscillation. An
epoch proposer assembles the synopses it has received, computes a new
\VortexHashName{} epoch, and circulates the compact model. Peers maintain one active
epoch on the data path and may stage one candidate epoch for rollout.

\paragraph{Epoch validity and activation.}
Epochs are scope-qualified and monotonically versioned. A candidate epoch
carries its scope, epoch number, compact model digest, region metadata version,
and expiration. Peers activate an epoch only after receiving the complete
compact state---centroids, region order, interval table, active-region set,
per-region model parameters, and metadata-version digest---and verifying its
digest. This rule  eventual convergence among live peers rather than consensus on the query path; lookup messages carry an epoch number,
and receivers reject requests for epochs that are neither active nor draining.
\VortexSystemName{} assumes authenticated control messages and benign crash failures; Byzantine fault tolerance is outside the current
model.

\paragraph{Epoch updates.}
\VortexSystemName{} does not stop the world when the learned hash changes. Epoch rollout is
two-phase. Once a candidate epoch has been fully received and validated, it
becomes the new \emph{write epoch}: new inserts and metadata updates use epoch
$e{+}1$ immediately. Existing vectors whose owners change under
$h_{\theta}^{(e+1)}$ are rehomed lazily in the background. During this drain,
epoch $e$ remains read-visible but read-only, and its region directory continues
to advertise draining owners. Queries route under the active epoch first. During rollover, they reserve a
bounded portion of the fanout budget for draining-epoch owners whose regions
overlap the active frontier, where overlap is determined by region identifier
or interval intersection with the active region hints. This keeps vectors not
yet rehomed read-visible without unbounded two-epoch fanout. Returned candidates
are deduplicated by vector identifier before final ranking. Once the drain
completes, the older epoch expires and the system returns to single-epoch
operation.

\paragraph{Routing states caching.}
Owner lookups and region-directory reads are hot control-path operations, so
ingress peers may cache two kinds of routing state. An \emph{owner hint} records
the scope, epoch, token range, and candidate owner that recently served a
learned-key lookup. A \emph{directory entry} records the scope, epoch, region
identifier, and peer set used for semantic expansion. Each cached record is
guarded by both a TTL and the epoch number. The TTL bounds staleness under
ordinary churn; the epoch guard prevents a peer from reusing state produced
under an older \VortexHashName{} ordering after the system switches epochs. During
rollover, cached entries for the active and draining epochs may coexist, but
epoch tags keep the views disjoint.

The cache is strictly advisory. A cache hit may shorten the common path: an
ingress peer can directly probe a hinted owner or reuse a cached region peer
set. The contacted peer still validates the scope, epoch, and token range or
region before serving the request. If an owner hint is stale, \VortexSystemName{} invalidates
it, may follow at most one bounded redirect, and then falls back to
authoritative \VortexOverlayName{} owner resolution. If the redirect target does not validate
the range, the ingress peer stops redirecting and performs authoritative \VortexOverlayName{}
lookup. If a directory entry is stale or missing, the ingress peer resolves the
live region peer set through \VortexOverlayName. \VortexSystemName{} therefore separates \emph{fast}
routing state from \emph{authoritative} routing state. The former may be cached
at the edge for latency, while the latter remains in the overlay and is
resolved through the same decentralized mechanism that governs ownership,
repair, and region metadata under churn and epoch rollover.


\subsection{\VortexHashName{} Model Selection}
\label{app:vortexhash-selection}

\VortexHashName{} is an epoch-versioned learned hash function,
$h_{\theta}^{(e)}:\mathbb{R}^{d}\rightarrow[0,2^b)$, whose output is the
ordered key used by \VortexOverlayName. Model selection is part of \VortexSystemName's maintenance loop,
not a one-time offline tuning step. While epoch $e$ serves queries, \VortexSystemName{}
compiles a candidate epoch $e{+}1$ off the query path from the current training
snapshot and the semantic synopses collected through heartbeats. A candidate
epoch is installed only after validation, so model training and selection do
not enter the query critical path.

Each candidate fixes a compact hash-model configuration: the synopsis budget
drawn from upper HNSW layers, the centroid codebook size, the centroid-ordering
policy, the density-aware interval allocator, the per-region CDF model family,
the CDF branching factor, and the 2-opt refinement budget for the
region order. Training a candidate produces a complete runtime hash state:
centroids define semantic regions, the learned region order places those
regions on the overlay ring, density-aware allocation assigns key-space
capacity, and per-region CDF models place vectors inside their assigned
intervals. The accepted epoch is emitted as a deterministic artifact containing
the model parameters, provenance, model size, and measured hash-inference
latency.

The selector optimizes \VortexHashName{} for its systems role: making decentralized
peer selection cheap while preserving retrieval quality. For a candidate
$\theta$, \VortexSystemName{} computes a placement score on held-out validation queries:
\[
\begin{aligned}
Q_{\mathrm{place}}(\theta)
&= \operatorname{clip}_{[0,1]}\!\Bigl( \\
&\quad 0.50\,M_{\mathrm{ovl}}(\theta)
+ 0.20\,M_{\mathrm{node}}(\theta) \\
&\quad + 0.15\,A_{\mathrm{win}}(\theta)
\quad + 0.10\,M_{\mathrm{tail}}(\theta) \\
&\quad + 0.05\,\bigl(1-D_{50}(\theta)\bigr)
\Bigr).
\end{aligned}
\]
All terms are normalized to $[0,1]$, with larger values better except
$D_{50}$. $M_{\mathrm{ovl}}$ measures whether owners of the exact nearest
neighbors appear in the bounded \VortexOverlayName{} window around $h_{\theta}(q)$.
$M_{\mathrm{node}}$ measures owner-level concentration of the ground-truth
answers. $A_{\mathrm{win}}$ is the area under the recall-versus-window curve as
the hash window expands. $M_{\mathrm{tail}}$ is fifth-percentile per-query
placement quality, which prevents a candidate from improving only easy queries.
$D_{50}$ is the median normalized hash-rank distance from the query key to a
responsible owner. Overlay match receives the largest weight because early
responsible-owner discovery is the signal that directly reduces fanout.

The placement-score weights are fixed across workloads and experiments. Each
component is computed on held-out validation queries for the current scope; test
queries are never used for epoch selection. The score is an engineering
selection rule for choosing a compact routing model, not a correctness proof.

Selection is constrained by deployment budgets:
\[
\theta^{*} =
\arg\max_{\theta \in \Theta}
Q_{\mathrm{place}}(\theta)
\quad
\text{s.t.}\quad
S(\theta)\le S_{\max},\;
L_{95}(\theta)\le L_{\max}.
\]
Here $S(\theta)$ is serialized model size and $L_{95}(\theta)$ is p95 hash
inference latency on validation queries. These constraints keep \VortexHashName{}
usable as hot-path routing logic rather than as a large auxiliary index.


\subsection{Experiments Details}
\label{app:workloads}

\subsubsection{Workloads and Ground Truth}
\label{app:workload-details}

Table~\ref{tab:app-workloads} summarizes the workloads and latency traces. All
vector workloads use L2 distance. Canonical base/query splits are used when
available. For slice-based workloads, ground truth is exact over the evaluated
slice. For scoped experiments, recall is computed against exact search over the
scope-eligible vector set, not against global nearest-neighbor labels. This is
necessary because scope is an eligibility constraint, not a post-filter.

\subsubsection{Scope Regimes}
\label{app:scope-construction}

Table~\ref{tab:app-scopes} summarizes the scope regimes. Uniform scopes weaken
the correlation between scope membership and embedding locality. Semantic scopes
model coherent workspaces. Mixed scopes model topical scoped collections with
outliers. Hot scopes stress serving load, and tiny scopes exercise the
small-scope fallback path.

\subsubsection{Shared Controls and Latency Regimes}
\label{app:shared-controls}

All methods use the same base/query split, exact ground truth, query stream,
ingress-peer distribution, sampled latency matrix, peer-local ANN backend,
local-search parameters, timeout policy, merge logic, and retrieval model.
Methods differ only in global placement, metadata maintenance, and peer
selection. Matched-recall experiments choose operating points on held-out
queries and evaluate those configurations on disjoint test queries. Hot-path
messages count query-caused routing, directory lookup, peer dispatch, and merge
messages; they exclude periodic maintenance, model dissemination, and
epoch-refresh traffic.

Table~\ref{tab:app-latency-regimes} lists the latency regimes used in
sensitivity experiments. Each regime preserves the relative heterogeneity of the
measured RTT matrix and changes only the absolute delay scale.

\subsubsection{Baseline Assumptions}
\label{app:baseline-assumptions}

Table~\ref{tab:app-baseline-assumptions} summarizes the routing and
coordination assumptions of each method. All methods use the same peer-local
ANN backend and merge path; the comparison isolates global placement, metadata,
and peer selection.

\subsection{Supplementary Results}
\label{app:evaluation}

\subsubsection{Recall Tail and Retrieval-Cost Breakdown}
\label{app:tail}

Figure~\ref{fig:app-tail-retrieval} expands the evaluation result by
breaking down where latency and query-path cost are spent.

Panel~(a) decomposes p95 end-to-end latency into routing, peer-local ANN,
merge, and retrieval fetch. Panel~(b) shows the corresponding query-path cost. Panel~(c) reports retrieval-deadline success at the matched-recall operating
point. The breakdown supports the main systems claim: \VortexSystemName's learned semantic frontier reduces query-path work and keeps
retrieval latency within the deadline while preserving decentralized ownership
and lookup.

\subsubsection{Scale-Up on the SPACEV Workload}
\label{app:scale-details}

We evaluate \VortexSystemName{} at larger scale using SPACEV1B in a controlled
1{,}024-node simulation. The experiment contains ten scopes, each with
10\,M eligible vectors and 256 hosting nodes on average. For each routing
budget, we compute recall@10 and p95 end-to-end retrieval latency per scope and
report the mean across the ten scopes. Figure~\ref{fig:app-scale-spacev} shows that \VortexSystemName{} preserves its recall-cost
behavior at scale.

\subsection{Artifact and Reproducibility}
\label{app:artifact}

The artifact is organized across three open-source projects.
\emph{LearnedHash} contains the \VortexHashName{} implementation and model-selection
code~\cite{VortexHashRepo}. \emph{Vortex System} contains the distributed
retrieval system, baseline implementations, experiment configurations, and
plotting scripts~\cite{VortexSystemRepo}. \emph{OverlaySim} contains the
event-driven substrate used for controlled scale, latency, churn, and refresh
experiments~\cite{OverlaySimRepo}.

\begin{table*}[p]
\centering

\begingroup
\footnotesize
\setlength{\tabcolsep}{4pt}
\renewcommand{\arraystretch}{0.96}
\caption{Evaluation workloads and latency traces.}
\label{tab:app-workloads}
\begin{tabularx}{\textwidth}{@{}lcccl>{\raggedright\arraybackslash}X@{}}
\toprule
Workload / trace & Size & Dim. & Queries / slices & Metric & Role \\
\midrule
SIFT1M~\cite{Aumuller2020ANNBenchmarks,TexMexCorpus}
  & 1M vectors & 128 & 10K queries & L2
  & standard ANN baseline \\
GIST1M~\cite{Aumuller2020ANNBenchmarks,TexMexCorpus}
  & 1M vectors & 960 & 1K queries & L2
  & high-dimensional descriptor baseline \\
SIFT10M~\cite{TexMexCorpus,Simhadri2022BigANN}
  & 10M vectors & 128 & 10K queries & L2
  & 10M descriptor stress test \\
DEEP10M-L2~\cite{Babenko2016Deep1B,Simhadri2022BigANN}
  & 10M vectors & 96 & 10K queries & L2
  & neural-embedding stress test \\
SPACEV10M~\cite{MicrosoftSPACEV1B,Simhadri2022BigANN}
  & 10M vectors & 100 & 29.3K queries & L2
  & production-style web-search embeddings \\
SPACEV1B~\cite{MicrosoftSPACEV1B,Simhadri2022BigANN}
  & 1B vectors & 100 & 29.3K queries & L2
  & billion-scale web-search peer-selection endpoint \\
\midrule
PlanetLab RTT~\cite{Zhu2017NetLatency,NetLatencyData}
  & 490 nodes & -- & 18 matrices & RTT
  & measured WAN latency input \\
\bottomrule
\end{tabularx}
\endgroup

\vspace{0.25\baselineskip}

\begingroup
\footnotesize
\setlength{\tabcolsep}{3pt}
\renewcommand{\arraystretch}{0.96}
\caption{Scope regimes.}
\label{tab:app-scopes}
\begin{tabularx}{\textwidth}{@{}l>{\raggedright\arraybackslash}p{0.18\textwidth}c>{\raggedright\arraybackslash}X>{\raggedright\arraybackslash}X@{}}
\toprule
Regime & Size distribution & Coherence & Construction & Purpose \\
\midrule
Uniform
& 10\% SPACEV10M scope
& Low
& Assign vectors uniformly at random while preserving scope size.
& Hard scoped case where eligibility and semantic locality are weakly aligned. \\
Semantic
& 10\% SPACEV10M scope
& High
& Form scopes from labels when available, otherwise from semantic regions.
& Project/workspace-like retrieval. \\
Mixed-Zipf
& 10\% SPACEV10M scope
& Medium
& Draw each scope from a Zipf-weighted mixture of semantic regions plus random
outliers.
& Topical collections with leakage and skew. \\
\bottomrule
\end{tabularx}
\endgroup

\vspace{0.25\baselineskip}

\begin{minipage}[t]{0.34\textwidth}
\centering
\scriptsize
\setlength{\tabcolsep}{2pt}
\renewcommand{\arraystretch}{0.96}
\caption{Latency regimes. }
\label{tab:app-latency-regimes}
\begin{tabularx}{\linewidth}{@{}>{\raggedright\arraybackslash}p{0.30\linewidth}>{\centering\arraybackslash}p{0.23\linewidth}>{\raggedright\arraybackslash}X@{}}
\toprule
Regime & Scale factor & Interpretation \\
\midrule
Cluster & 0.005 & datacenter or nearby-cluster deployment \\
Regional & 0.05 & nearby-site or edge-like deployment \\
Global-WAN & 1.0 & unscaled WAN stress case \\
\bottomrule
\end{tabularx}
\end{minipage}\hfill
\begin{minipage}[t]{0.63\textwidth}
\centering
\footnotesize
\setlength{\tabcolsep}{2pt}
\renewcommand{\arraystretch}{0.96}
\caption{Baselines.}
\label{tab:app-baseline-assumptions}
\begin{tabularx}{\linewidth}{@{}>{\raggedright\arraybackslash}p{0.21\linewidth}>{\raggedright\arraybackslash}p{0.18\linewidth}>{\centering\arraybackslash}p{0.07\linewidth}>{\raggedright\arraybackslash}X>{\centering\arraybackslash}p{0.09\linewidth}>{\centering\arraybackslash}p{0.11\linewidth}@{}}
\toprule
Method & Global routing state & {\scriptsize Query router} & Semantic mechanism &
{\scriptsize Same local ANN} & {\scriptsize Weakly coordinated} \\
\midrule
\textsc{RandomSelector}
& none
& no
& none; exact-key ring exploration
& yes
& yes \\
\textsc{DET-LSH}
& LSH bucket metadata
& no
& decentralized LSH buckets and multiprobe lookup
& yes
& yes \\
\textsc{Distributed ScaNN}
& partition metadata at routing tier
& yes
& hierarchical partition routing
& yes
& partial \\
\textsc{Centralized Router}
& fresh global routing metadata
& yes
& centralized semantic route selection
& yes
& no \\
\textsc{\VortexSystemName}
& compact \VortexHashName{} state and \VortexOverlayName{} metadata
& no
& learned ordered keyspace with bounded repair
& yes
& yes \\
\bottomrule
\end{tabularx}
\end{minipage}
\end{table*}

\begin{figure*}[p]
\begin{minipage}[t]{0.65\textwidth}
  \centering
  \includegraphics[width=\linewidth]{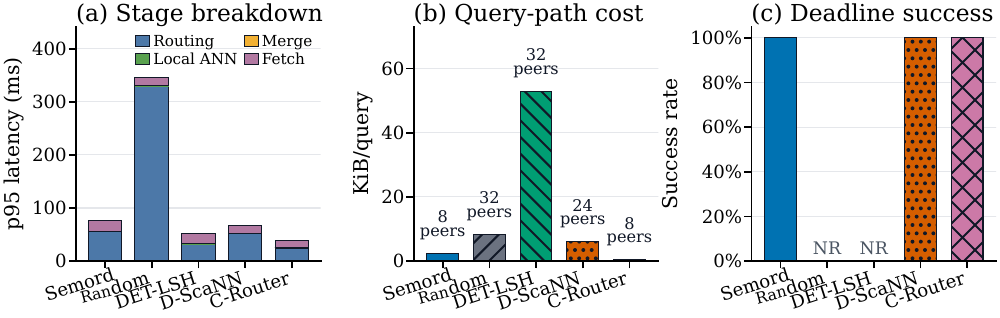}
  \Description{Matched-recall retrieval-cost figure showing latency breakdown,
  query-path bytes, and retrieval-deadline success at matched recall.}
  \VortexTighten{-2ex}
  \caption{\textbf{Retrieval cost.}}
  \label{fig:app-tail-retrieval}
\end{minipage}\hfill
\begin{minipage}[t]{0.32\textwidth}
  \centering
  \includegraphics[width=\linewidth]{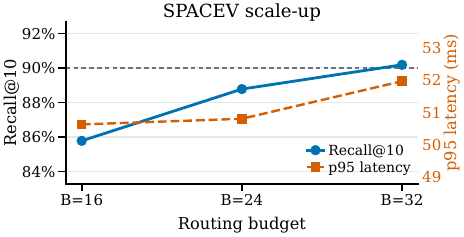}
  \Description{SPACEV1B scale-up figure showing recall-cost behavior at larger
  scale across routing budgets.}
  \VortexTighten{-2ex}
  \caption{\textbf{SPACEV1B scale-up.}}
  \label{fig:app-scale-spacev}
\end{minipage}
\end{figure*}

\fi

\end{document}